\documentclass[usenatbib,useAMS]{mn2e}
\usepackage{epsfig}
\usepackage{mathptmx}
\usepackage{amsmath}

\usepackage{verbatim}

\newcommand{\simgt}%
        {\,\hbox{\lower0.6ex\hbox{$\sim$}\llap{\raise0.6ex\hbox{$>$}}}\,}
\newcommand{\simlt}%
        {\,\hbox{\lower0.6ex\hbox{$\sim$}\llap{\raise0.6ex\hbox{$<$}}}\,}
\usepackage{color}
\newcommand{\msun}{\ensuremath{\mathrm{M}_\odot}}

\title[]{Synthetic light curves and spectra for three-dimensional delayed-detonation models of Type Ia supernovae}
\author[Sim et al.]
{\parbox{\textwidth}{S. A. Sim$^{1,2,3}$, I. R. Seitenzahl$^{4,5}$, M. Kromer$^{5}$, F. Ciaraldi-Schoolmann$^5$,  F. K. R\"{o}pke$^{4}$, 
M. Fink$^{4}$, W. Hillebrandt$^5$, R. Pakmor$^6$, A. J. Ruiter$^5$, S. Taubenberger$^5$}\vspace{0.4cm}\\
$^{1}$Astrophysics Research Centre, School of Mathematics and Physics, Queen's University Belfast, Belfast BT7 1NN, UK\\
$^{2}$Research School of Astronomy and Astrophysics, Mount Stromlo Observatory,
Cotter Road, Weston Creek, ACT 2611, Australia\\ 
$^{3}$ARC Centre of Excellence for All-sky Astrophysics (CAASTRO)\\
$^{4}$Universit{\"a}t W{\"u}rzburg, Emil-Fischer-Str.~31, D-97074 W{\"u}rzburg, Germany\\
$^{5}$Max-Planck-Institut f\"{u}r Astrophysik, 
Karl-Schwarzschildstr. 1, D-85748 Garching, Germany\\
$^{6}$Heidelberger Institut f\"{u}r Theoretische Studien, Schloss-Wolfsbrunnenweg 35, D-69118 Heidelberg, Germany
}

\date{Accepted 2013 August 19.  Received 2013 August 2; in original form 2013 April 23}

\pagerange{\pageref{firstpage}--\pageref{lastpage}}
\pubyear{2013}
\begin{document}
\maketitle
\label{firstpage}

\begin{abstract}
In a companion paper, Seitenzahl et al. (2013) have presented a set of three-dimensional delayed detonation models for thermonuclear explosions of near-Chandrasekhar mass white dwarfs (WDs).
Here, we present multi-dimensional radiative transfer simulations that provide synthetic light curves and spectra for those models.
The model sequence explores both changes in the strength of the deflagration phase (which is controlled by the ignition configuration in our models) and the WD central density. In agreement with previous studies, we find that the strength of the deflagration significantly affects the explosion and the observables. Variations in the central density also have an influence on both brightness and colour, but overall it is a secondary parameter in our set of models.
In many respects, the models yield a good match to the observed properties of normal Type~Ia supernovae (SNe~Ia): peak brightness, rise/decline time scales and synthetic spectra are all in reasonable agreement. 
There are, however, several differences. 
In particular, the models are systematically too red around maximum light, manifest spectral line velocities that are a little too high and yield $I$-band light curves that do not match observations. 
Although some of these discrepancies may simply relate to approximations made in the modelling, 
some pose real challenges to the models.
If viewed as a complete sequence, our models do not reproduce the observed light-curve width-luminosity relation (WLR) of SNe~Ia: all our models show rather similar $B$-band decline rates, irrespective of peak brightness. 
This suggests that simple variations in 
the strength of the deflagration phase
in Chandrasekhar-mass deflagration-to-detonation models do not readily explain the observed diversity of normal SNe~Ia.
This may imply that some other parameter within the Chandrasekhar-mass paradigm is key to the WLR, or that a substantial fraction of normal SNe~Ia arise from an alternative explosion scenario.
\end{abstract}


\section{Introduction}
\label{sect_intro}

It is generally accepted that Type~Ia supernovae (SNe~Ia) are the result of the thermonuclear explosion of carbon-oxygen (CO) white dwarf (WD) stars \citep[see e.g.\ ][]{hillebrandt00,hillebrandt13}. However, the evolutionary channel leading to explosion and the mechanisms by which the thermonuclear flame ignites and propagates remain unclear.
The best known and most widely studied explosion model for SNe~Ia is the single-degenerate Chandrasekhar-mass scenario. In that model, a CO WD gains mass by accretion from a non-degenerate companion star. When the WD mass draws near to the Chandrasekhar limit, carbon burning ignites near to the centre of the WD. Although initially neutrino cooling is sufficient to prevent explosion, heat released by carbon burning eventually gives rise to local conditions where a thermonuclear runaway can occur, leading to the birth of a thermonuclear deflagration flame.

Three-dimensional (3D) explosion simulations of pure deflagration models for SNe~Ia have been considered in several studies 
\citep[e.g.\ ][]{reinecke02,gamezo03,roepke05,roepke08}.
Although it is possible for deflagration models to yield explosions in which the WD is entirely disrupted, such models struggle to produce a sufficiently large mass of $^{56}$Ni to account for the observed brightness of normal SNe~Ia: even with favourably chosen initial conditions, only around ${\sim}0.33$~M$_{\odot}$ of $^{56}$Ni are produced (\citealt{roepke07c}; see also \citealt{ma13})\footnote{In addition, \cite{kozma05} found that the degree of mixing in deflagration models should result in significant O emission at late times, which is inconsistent with the observed nebular-phase spectra of normal SNe~Ia.}. Deflagration models are therefore most promising for explaining certain sub-luminous SNe~Ia. We also note that, depending on the ignition configuration, deflagration simulations can give rise to explosions in which only a part of the mass of the WD is ejected \citep{jordan12b} -- this scenario gives a good match to the observed properties of the faint/spectroscopically peculiar class of 2002cx-like SNe~Ia \citep{kromer13}.

To account for the brightness of normal SNe~Ia, more $^{56}$Ni must be synthesised than is expected from pure deflagration models. For near-Chandrasekhar mass models, this can be achieved by invoking a thermonuclear detonation at some point during or after the propagation of the deflagration flame. A few such ``delayed detonation'' mechanisms have been proposed, including spontaneous deflagration-to-detonation transition (DDT) models \citep{khokhlov91}, the gravitationally confined detonation model \citep{plewa04},
the pulsational detonation model \citep[e.g.\ ][]{khokhlov92,bravo09a,bravo09b}
and the pulsationally assisted gravitationally confined detonation model \citep{jordan12a}. In this study, we will focus on DDT models, in which it is argued that the deflagration flame transitions into a detonation during the late stages of the explosion. This may occur when, due to the low fuel densities, the flame structure broadens and turbulence starts to affect the microphysical processes. If in this ``distributed burning regime'' turbulent velocity fluctuations are strong enough to mix fuel and ashes efficiently, a detonation may ignite via the Zel'dovich gradient mechanism 
\citep[e.g.,][]{woosley07,woosley09}.

Synthetic spectra and light curves have been presented for sets of
near-Chandrasekhar mass DDT models in several previous studies. One-dimensional (1D) DDT models have been studied by e.g.\ \cite{hoeflich95}, \cite{hoeflich96a} and \cite{blondin13}, while 2D simulations were presented by \citet{kasen09} and \citet{blondin11}. Dimensionality, however, is an important consideration for DDT models. Since the propagation of the deflagration flame is controlled by turbulence and the hot deflagration ash is subject to buoyancy (and secondary) instabilities, full 3D simulations are needed both to model the flame propagation and predict the final composition of the ejecta. Recently,  
\cite{seitenzahl13} presented a set of fourteen 3D DDT explosion simulations (and associated nucleosynthesis post-processing) for SNe~Ia. These models differ from each other in the adopted ignition configuration of the deflagration flame and in the central density of the WD. Therefore, they provide a useful set of models with which to consider whether fully 3D DDT models can account for the observed properties of normal SNe~Ia and to investigate how observable quantities are affected by key initial conditions of explosion simulations. To this end, in this paper we 
present the results of radiative transfer calculations for the \cite{seitenzahl13} models and compare them with observed properties of SNe~Ia.

In Section~\ref{sect:sims} we remind the reader of salient details of the \cite{seitenzahl13} models and describe our radiative transfer calculations. We present results of our calculations in Section~\ref{sect:results}. Implications of our work are discussed in Section~\ref{sect:discuss} and conclusions are summarised in Section~\ref{sect:summary}.

\section{Simulations}
\label{sect:sims}

\subsection{Explosion models}
\label{sect:models}

Below we will present the results of radiative transfer calculations
applied to the set of fourteen explosion models described by
\cite{seitenzahl13}. The hydrodynamics simulations were carried out
using {\sc leafs}, which is a 3D finite-volume code that uses a
levelset method to track thermonuclear flame fronts \cite[see][and references therein]{seitenzahl13}.
As initial conditions for the explosion simulation,  \cite{seitenzahl13} adopted a cold hydrostatic WD with uniform composition: $^{12}$C/$^{16}$O/$^{22}$Ne of 47.5/50/2.5 per cent by mass. In a future study, we plan to investigate the consequences of adopting more realistic, non-uniform initial compositions, but for all models in this paper, the WD composition is homogeneous.

The explosions are ignited by inserting a distribution of deflagration sparks: small spherical ($10^6$~cm radius) regions where it is assumed that a local thermonuclear runaway has given birth to a deflagration. 
It is well established from multi-dimensional simulations that the ignition configuration can significantly affect the outcome of DDT models \citep{golombek05,roepke07,mazzali07,bravo08,kasen09,krueger12,seitenzahl13}.
In our models, a larger number of sparks means that the total surface area of the deflagration is initially bigger, leading to more rapid fuel consumption. Also, more sparks lead to higher order spatial modes being excited in the structure of the flame front, affecting the growth of instabilities and the development of turbulence. It follows that models ignited with large numbers of sparks have {\it stronger} deflagration phases\footnote{Throughout this paper, by ``stronger'' deflagration we mean that the rate of fuel consumption during the deflagration phase is larger. This leads to more energy release, and consequently more expansion of the WD, prior to the detonation.}. 
We stress that the number of ignition sparks in any given model should not be interpreted literally: the numbers of sparks used in this study were chosen, by trial and error, in order to produce models with a sufficiently wide range of $^{56}$Ni masses to encompass the bulk of normal SNe~Ia. Thus, the models are not motivated by consideration of ignition conditions and do not address the question of whether deflagrations of the corresponding strength would be realised in Nature. Instead, the purpose of these models is twofold. First, to investigate whether a good match to observed SNe~Ia can be found for some choice of the deflagration strength. Second, to explore 
the hypothesis that simple variations in the deflagration strength in DDT models might explain the observed diversity of SNe~Ia. In this context, the number / location of sparks is merely a convenient parametrization that allows us to realise a set of models 
with a wide range of rates of fuel consumption during the deflagration phase. 

Twelve of the fourteen models form a sequence along which the models differ {\it only} in the number (and location) of the ignition sparks adopted for the deflagration flame. 
In these models, the number of ignition sparks varies from 1 to 1600: throughout this paper we will refer to the individual models by their number of ignition sparks as Nx (x ranging from 1 to 1600). 
The ignition sparks are placed fairly close to the centre of the WD ($r < 2.5 \times 10^7$~cm). In two of the models the sparks are particularly centrally concentrated: these two models are identified by NxC. For details of the ignition geometries and the algorithm used to place the sparks, see \cite{seitenzahl13}. In each of these twelve models, the WD central density was chosen to be $2.9 \times 10^9$~g~cm$^{-3}$ (which gives a total WD mass of $1.400$~\msun).

The remaining two models differ from the other twelve in the choice of WD central density. In one case a low central density of $1 \times 10^9$~g~cm$^{-3}$ ($1.361$~\msun) is adopted, while in the other a high value of $5.5 \times 10^9$~g~cm$^{-3}$ ($1.416$~\msun) is used. 
The WD central density has been suggested as a potential secondary parameter that can affect the outcome of a DDT explosion, primarily via its influence on the masses of both $^{56}$Ni and stable iron-group elements that are synthesised during nuclear burning at high density \citep{krueger11,seitenzahl11,krueger12,seitenzahl13}\footnote{We note, however, that the exact role of central density is dependent on the explosion modelling: e.g.\ \cite{krueger12} predict that the (mean) effect of increased central density is to {\it reduce} the $^{56}$Ni yield (due to more neutronization via electron captures) while in our models \citep{seitenzahl13}, the $^{56}$Ni yield {\it increases} since the rate of increase in total production of iron-group elements outpaces the rise in stable iron-group elements \citep[see discussions in ][]{seitenzahl11,krueger12,seitenzahl13}.}.
To facilitate direct comparison, the ignition geometries for our two models with different central densities were chosen to be the same as for the N100 model. We will therefore refer to the models with high and low central densities as N100H and N100L, respectively.

\begin{table*}
\caption{Asymptotic ejecta kinetic energies, $^{56}$Ni masses, light-curve rise and decline parameters, peak absolute optical magnitudes and colours for our simulations. For each quantity, the value derived from the angle-averaged light curve is quoted along with the range obtained from our
full set of orientation-dependent light curves (specified as a $\pm$ range on each derived value).}
\begin{tabular}{lcccccccccc}\hline
Model &  $E_{\rm kin}$ &$M(^{56}$Ni$)$& $t_{\rm max}^{\rm bol}$  &$t_{\rm max}^B$ & $M^{\rm bol}_{\rm max}$ &$U_{\rm max}$ & $B_{\rm max}$ & $V_{\rm max}$& $R_{\rm max}$ & $I_{\rm max}$\\
 & $10^{51}$~erg & M$_{\odot}$ & days$^a$ & days & mag$^a$ & mag & mag & mag & mag & mag\\
\hline
N1 & $1.67$ & $1.10$ & $16.6_{-0.4}^{+0.4}$ & $16.5_{-0.6}^{+0.4}$ & $-19.48^{-0.06}_{+0.06}$ & $-20.39^{-0.13}_{+0.17}$ & $-19.93^{-0.06}_{+0.05}$ & $-20.12^{-0.03}_{+0.02}$ & $-19.56^{-0.05}_{+0.04}$ & $-19.70^{-0.05}_{+0.05}$\\[0.2cm]
N3 & $1.61$ & $1.03$ & $17.7_{-0.6}^{+1.2}$ & $17.4_{-0.7}^{+0.6}$ & $-19.40^{-0.21}_{+0.40}$ & $-20.18^{-0.40}_{+1.19}$ & $-19.80^{-0.22}_{+0.57}$ & $-20.08^{-0.11}_{+0.13}$ & $-19.59^{-0.09}_{+0.06}$ & $-19.63^{-0.13}_{+0.14}$\\[0.2cm]
N5 & $1.59$ & $0.97$ & $18.0_{-0.7}^{+0.9}$ & $17.7_{-0.7}^{+0.9}$ & $-19.33^{-0.16}_{+0.21}$ & $-20.05^{-0.31}_{+0.50}$ & $-19.71^{-0.18}_{+0.31}$ & $-20.00^{-0.09}_{+0.09}$ & $-19.60^{-0.08}_{+0.06}$ & $-19.60^{-0.10}_{+0.09}$\\[0.2cm]
N10 & $1.60$ & $0.93$ & $18.1_{-0.7}^{+0.7}$ & $17.7_{-0.5}^{+0.5}$ & $-19.29^{-0.17}_{+0.11}$ & $-19.94^{-0.30}_{+0.25}$ & $-19.66^{-0.16}_{+0.14}$ & $-19.96^{-0.10}_{+0.06}$ & $-19.60^{-0.12}_{+0.07}$ & $-19.59^{-0.11}_{+0.06}$\\[0.2cm]
N20 & $1.52$ & $0.77$ & $18.3_{-1.1}^{+1.2}$ & $17.7_{-1.0}^{+1.0}$ & $-19.09^{-0.13}_{+0.07}$ & $-19.55^{-0.26}_{+0.15}$ & $-19.34^{-0.15}_{+0.16}$ & $-19.73^{-0.07}_{+0.06}$ & $-19.55^{-0.09}_{+0.06}$ & $-19.51^{-0.11}_{+0.08}$\\[0.2cm]
N100H & $1.53$ & $0.69$ & $18.2_{-0.9}^{+1.0}$ & $16.8_{-1.1}^{+1.0}$ & $-18.96^{-0.09}_{+0.08}$ & $-19.08^{-0.23}_{+0.22}$ & $-19.10^{-0.18}_{+0.14}$ & $-19.65^{-0.08}_{+0.07}$ & $-19.57^{-0.04}_{+0.04}$ & $-19.53^{-0.05}_{+0.08}$\\[0.2cm]
N40 & $1.47$ & $0.65$ & $18.3_{-1.1}^{+0.7}$ & $17.3_{-1.0}^{+0.7}$ & $-18.93^{-0.07}_{+0.09}$ & $-19.14^{-0.28}_{+0.33}$ & $-19.11^{-0.12}_{+0.16}$ & $-19.57^{-0.05}_{+0.07}$ & $-19.50^{-0.03}_{+0.03}$ & $-19.46^{-0.08}_{+0.06}$\\[0.2cm]
N100 & $1.46$ & $0.60$ & $18.2_{-0.7}^{+0.8}$ & $17.0_{-0.7}^{+0.7}$ & $-18.87^{-0.07}_{+0.05}$ & $-18.98^{-0.19}_{+0.22}$ & $-19.02^{-0.10}_{+0.07}$ & $-19.51^{-0.05}_{+0.06}$ & $-19.47^{-0.03}_{+0.03}$ & $-19.44^{-0.04}_{+0.04}$\\[0.2cm]
N150 & $1.41$ & $0.56$ & $18.0_{-0.6}^{+0.9}$ & $16.5_{-0.9}^{+1.0}$ & $-18.80^{-0.05}_{+0.08}$ & $-18.94^{-0.14}_{+0.18}$ & $-18.82^{-0.10}_{+0.13}$ & $-19.39^{-0.05}_{+0.05}$ & $-19.43^{-0.03}_{+0.04}$ & $-19.43^{-0.04}_{+0.05}$\\[0.2cm]
N100L & $1.37$ & $0.53$ & $18.2_{-1.1}^{+0.9}$ & $17.2_{-1.0}^{+0.8}$ & $-18.76^{-0.09}_{+0.12}$ & $-18.89^{-0.30}_{+0.37}$ & $-18.89^{-0.16}_{+0.22}$ & $-19.34^{-0.09}_{+0.10}$ & $-19.34^{-0.05}_{+0.04}$ & $-19.29^{-0.06}_{+0.07}$\\[0.2cm]
N300C & $1.42$ & $0.51$ & $17.4_{-0.8}^{+0.7}$ & $16.0_{-1.4}^{+0.8}$ & $-18.76^{-0.04}_{+0.06}$ & $-18.70^{-0.13}_{+0.14}$ & $-18.69^{-0.11}_{+0.17}$ & $-19.29^{-0.06}_{+0.08}$ & $-19.38^{-0.04}_{+0.05}$ & $-19.41^{-0.06}_{+0.06}$\\[0.2cm]
N200 & $1.34$ & $0.41$ & $18.1_{-1.3}^{+0.9}$ & $16.1_{-2.0}^{+1.2}$ & $-18.53^{-0.07}_{+0.11}$ & $-18.46^{-0.18}_{+0.25}$ & $-18.46^{-0.12}_{+0.20}$ & $-19.06^{-0.06}_{+0.13}$ & $-19.20^{-0.06}_{+0.10}$ & $-19.28^{-0.06}_{+0.05}$\\[0.2cm]
N1600 & $1.32$ & $0.36$ & $18.4_{-0.8}^{+1.0}$ & $16.0_{-1.1}^{+1.4}$ & $-18.39^{-0.07}_{+0.09}$ & $-18.10^{-0.26}_{+0.30}$ & $-18.26^{-0.14}_{+0.21}$ & $-18.93^{-0.06}_{+0.10}$ & $-19.10^{-0.05}_{+0.09}$ & $-19.18^{-0.05}_{+0.04}$\\[0.2cm]
N1600C & $1.20$ & $0.32$ & $18.5_{-0.7}^{+0.6}$ & $16.0_{-0.7}^{+0.6}$ & $-18.26^{-0.06}_{+0.08}$ & $-18.13^{-0.36}_{+0.29}$ & $-18.16^{-0.11}_{+0.14}$ & $-18.73^{-0.06}_{+0.07}$ & $-18.94^{-0.03}_{+0.05}$ & $-19.06^{-0.06}_{+0.05}$ \\\hline
\end{tabular}\\

\begin{tabular}{lccccc}\hline
Model &  $B - V$ & $V - R$& $V - I$& $\Delta m_{15}^B$& $\Delta m_{15}^V$\\
 & mag$^b$ & mag$^b$ & mag$^b$ & mag & mag \\
\hline
N1 & $  0.15_{-0.04}^{+0.06}$ & $ -0.53_{-0.05}^{+0.04}$ & $ -1.24_{-0.05}^{+0.05}$ & $  1.36_{-0.11}^{+0.11}$ & $  0.70_{-0.05}^{+0.04}$\\[0.2cm]
N3 & $  0.23_{-0.12}^{+0.44}$ & $ -0.45_{-0.07}^{+0.10}$ & $ -1.11_{-0.13}^{+0.15}$ & $  1.27_{-0.36}^{+0.20}$ & $  0.74_{-0.08}^{+0.08}$\\[0.2cm]
N5 & $  0.23_{-0.10}^{+0.25}$ & $ -0.36_{-0.07}^{+0.06}$ & $ -0.93_{-0.08}^{+0.09}$ & $  1.34_{-0.25}^{+0.13}$ & $  0.77_{-0.08}^{+0.11}$\\[0.2cm]
N10 & $  0.24_{-0.07}^{+0.08}$ & $ -0.31_{-0.04}^{+0.04}$ & $ -0.82_{-0.08}^{+0.09}$ & $  1.37_{-0.13}^{+0.10}$ & $  0.79_{-0.06}^{+0.10}$\\[0.2cm]
N20 & $  0.31_{-0.10}^{+0.11}$ & $ -0.13_{-0.06}^{+0.08}$ & $ -0.40_{-0.12}^{+0.12}$ & $  1.34_{-0.17}^{+0.09}$ & $  0.83_{-0.11}^{+0.05}$\\[0.2cm]
N100H & $  0.43_{-0.11}^{+0.09}$ & $ -0.03_{-0.05}^{+0.05}$ & $ -0.20_{-0.08}^{+0.10}$ & $  1.33_{-0.12}^{+0.13}$ & $  0.86_{-0.06}^{+0.08}$\\[0.2cm]
N40 & $  0.35_{-0.07}^{+0.10}$ & $ -0.02_{-0.05}^{+0.05}$ & $ -0.16_{-0.10}^{+0.08}$ & $  1.36_{-0.08}^{+0.08}$ & $  0.87_{-0.08}^{+0.04}$\\[0.2cm]
N100 & $  0.37_{-0.07}^{+0.07}$ & $  0.02_{-0.04}^{+0.04}$ & $ -0.09_{-0.08}^{+0.07}$ & $  1.39_{-0.08}^{+0.09}$ & $  0.88_{-0.09}^{+0.06}$\\[0.2cm]
N150 & $  0.48_{-0.07}^{+0.07}$ & $  0.07_{-0.04}^{+0.05}$ & $  0.02_{-0.05}^{+0.06}$ & $  1.30_{-0.13}^{+0.10}$ & $  0.85_{-0.10}^{+0.06}$\\[0.2cm]
N100L & $  0.35_{-0.11}^{+0.11}$ & $  0.05_{-0.05}^{+0.07}$ & $ -0.03_{-0.12}^{+0.08}$ & $  1.34_{-0.16}^{+0.12}$ & $  0.85_{-0.11}^{+0.09}$\\[0.2cm]
N300C & $  0.49_{-0.09}^{+0.08}$ & $  0.11_{-0.04}^{+0.06}$ & $  0.11_{-0.06}^{+0.10}$ & $  1.29_{-0.22}^{+0.11}$ & $  0.86_{-0.11}^{+0.10}$\\[0.2cm]
N200 & $  0.50_{-0.06}^{+0.07}$ & $  0.15_{-0.03}^{+0.05}$ & $  0.19_{-0.09}^{+0.11}$ & $  1.26_{-0.27}^{+0.14}$ & $  0.88_{-0.14}^{+0.07}$\\[0.2cm]
N1600 & $  0.56_{-0.11}^{+0.12}$ & $  0.19_{-0.04}^{+0.04}$ & $  0.24_{-0.07}^{+0.10}$ & $  1.16_{-0.19}^{+0.14}$ & $  0.88_{-0.08}^{+0.08}$\\[0.2cm]
N1600C & $  0.46_{-0.07}^{+0.07}$ & $  0.20_{-0.04}^{+0.04}$ & $  0.28_{-0.06}^{+0.07}$ & $  1.17_{-0.14}^{+0.13}$ & $  0.79_{-0.08}^{+0.10}$\\ \hline
\end{tabular}\\ 
$^a$ The rise time $t_{\rm max}^{\rm bol}$ and peak absolute magnitude $M^{\rm bol}_{\rm max}$ are derived from our $UBVRIJHK$ bolometric light curves. \\
$^b$ Colours are given at the epoch of $B$-band maximum light.
\label{tab:lcs}
\end{table*}

For all the simulations in this study, the same DDT criterion was used, based on the strength of turbulent velocity fluctuations at the flame front 
(see \citealt{seitenzahl13} and \citealt{ciaraldi13} for details). In all models at least one detonation occurred and the WD was always completely disrupted by the release of thermonuclear energy. By design (see above), the model sequence explores a sufficiently wide range of deflagration strengths that 
the range of $^{56}$Ni masses predicted by the simulations is large (0.32 -- 1.1~M$_{\odot}$). The $^{56}$Ni mass is generally anti-correlated with the number of ignition sparks adopted (i.e., a larger number of ignition sparks leads to a stronger deflagration phase and more expansion of the WD before detonation occurs).
Table~{\ref{tab:lcs}} lists the models in order of synthesised $^{56}$Ni mass and gives both the total $^{56}$Ni mass and the asymptotic explosion energy ($E_{\rm kin}$) in each case.

The {\sc leafs} simulations were run until $t = 100$~s after ignition, by which point the dynamics of the ejecta were very close to free expansion. 
\cite{seitenzahl13} carried out post-processing nucleosynthesis calculations (384-isotope network) to obtain detailed nucleosynthesis yields for each model. Those calculations were performed using the thermodynamic trajectories of $10^6$ passive Lagrangian tracer particles that were recorded during the hydrodynamics simulations \citep[following][]{travaglio04,seitenzahl10a}. We have used the final positions (and compositions) of the tracer particles to reconstruct the full 3D composition of the ejecta via a smoothed-particle-hydrodynamics-like algorithm \citep[as described by][]{kromer10}.

\subsection{Radiative transfer simulations}

For each model, we use the {\sc artis} code \citep{sim07,kromer09} to compute synthetic light curves and spectra. {\sc artis} employs a Monte Carlo (MC) algorithm \citep{lucy02,lucy03,lucy05} to carry out multi-dimensional, time-dependent radiative transfer calculations for homologously expanding SN ejecta. 

As input to {\sc artis}, we take the density and composition structure obtained from the final state of the explosion simulations (Section~\ref{sect:models}), remapped to a uniform 3D Cartesian grid of $50^3$ cells.
For all times after $t = 100$~s (the end of the {\sc leafs} simulations), the ejecta dynamics are assumed to be well-described by an homologous expansion law.
The {\sc artis} calculations were started at $t = 2$~days after explosion. As described by \cite{lucy05}, energy generated by radioactive decays before $t = 2$~days is included by assuming that it has been trapped and advected with the homologously expanding ejecta up to the start of the MC simulation.
The radiative transfer simulations are run until
$120$~days after explosion in a sequence of 111 logarithmically spaced time steps. For each simulation a total of $1.024 \times 10^8$ MC quanta were used. Angle-averaged spectra and light curves were obtained by appropriate binning of emergent MC quanta as functions of photon frequency and escape time. In addition, for each model, orientation-dependent synthetic observables were obtained by further binning of the quanta in 100 orientation bins (each of equal solid angle).

All radiative transfer calculations were performed using our atomic data set drawn from \cite{kurucz95} \citep[see][]{kromer09}, expanded to included ions {\sc i} -- {\sc vii} for $20 < Z < 29$, as in \cite{sim12}. We note, based on the comparisons in \cite{kromer09}, that the \cite{kurucz95} line list yields optical spectra that agree well with more expensive calculations involving an even larger atomic data set, making it adequate for this study (which will focus on optical comparisons). We note, however, that the near-infrared (NIR) region is more sensitive to the choice of atomic data 
[\cite{kromer09} 
found that using an extended set of atomic data for the iron-group elements led to an increase in the early time NIR light curves by several tenths of a magnitude, compared to our standard Kurucz \& Bell line list; see also \cite{kasen06b}]. This must be borne in mind when comparing our synthetic NIR light curves to observations.

\section{Results}
\label{sect:results}

\subsection{Synthetic light curves}
\label{sect:lcs}

In this section, we will focus on presenting the synthetic light curves and colours of our models, deferring discussion of spectral features to the next section.

\subsubsection{Angle-averaged light curves}
\begin{figure*}
  \centering
  \includegraphics[height=16cm, angle=90]{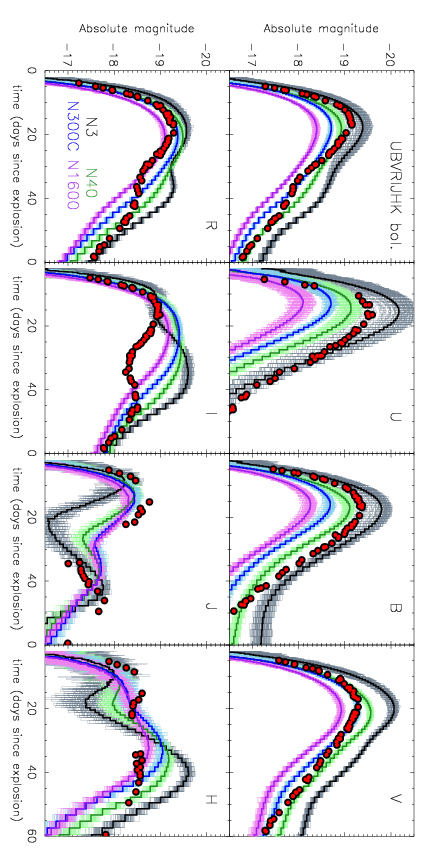}
\caption{Angle-averaged light curves ($UBVRIJHK$ bolometric, $U$-, $B$-, $V$-, $R$-, $I$-, $J$-, and $H$-bands) for four of our explosion models with differing deflagration strengths (heavy-drawn lines). To indicate the degree of sensitivity to orientation, the corresponding bands of lightly drawn lines shown the light curves from each of our 100 viewing angle bins. Note that Monte Carlo noise affects the angle-dependent light curves, particularly in bands/at epochs when the flux is relatively small (e.g.\ the $H$-band at early times). In each panel we show observations of SN~2005cf (filled circles) from Pastorello et al. (2007). For ease of comparison, the observed light curves are plotted assuming a $B$-band rise time of 17~days and are corrected for reddening (using parameters from Pastorello et al. 2007).}
\label{fig:lcs_1}
\end{figure*}

\begin{figure*}
\includegraphics[height=16cm, angle=90]{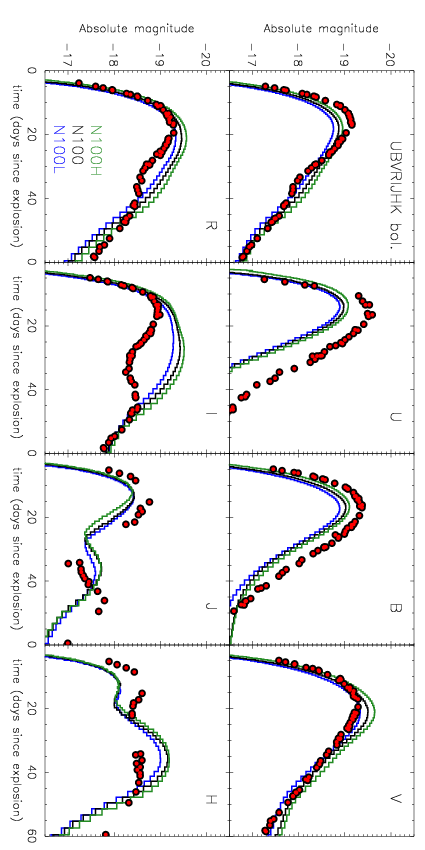}
\caption{As Figure 1, but showing our explosion models with differing WD central densities. For clarity, angle-dependent light curves are omitted here.}
\label{fig:lcs_2}
\end{figure*}

Bolometric ($UBVRIJHK$) and $B$-band rise times ($t_{\rm max}^{\rm bol}$ and $t_{\rm max}^B$, respectively), peak absolute magnitudes, optical colours at $t_{\rm max}^B$ and values of the decline rate parameters\footnote{$\Delta m_{15}^X$ is defined as the increase in $X$-band magnitude during a 15~day period after maximum light in the $X$-band.} $\Delta m_{15}^B$ and $\Delta m_{15}^V$ derived from our angle-averaged light curves are tabulated in Table~{\ref{tab:lcs}}. 
Complete synthetic angle-averaged bolometric ($UBVRIJHK$) and photometric band-limited light curves are shown for subsets\footnote{Specifically, the figures show the light curves for the models shown in figures 3 and 4 of Seitenzahl et al. (2013).} of our models in Figures~\ref{fig:lcs_1} and \ref{fig:lcs_2}.

\begin{figure}
\includegraphics[height=8cm, angle=90]{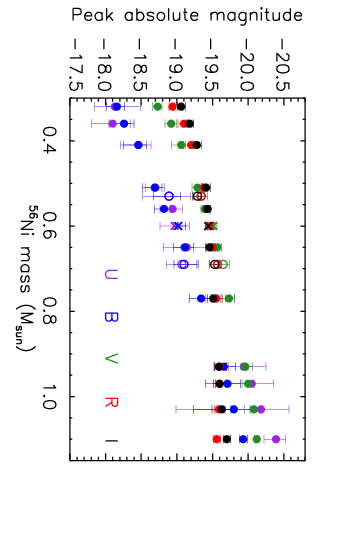}
\includegraphics[height=8cm, angle=90]{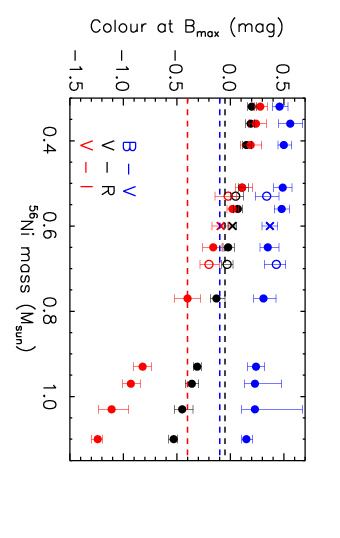}
\includegraphics[height=8cm, angle=90]{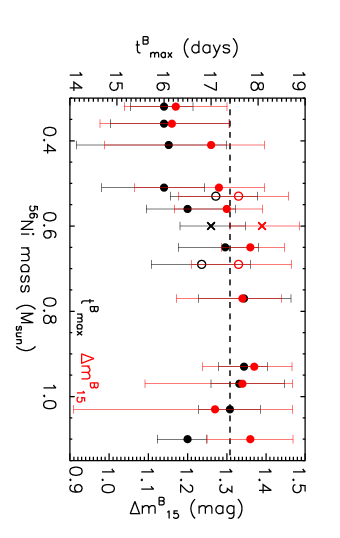}
\caption{Light curve properties of the models versus $^{56}$Ni mass. Filled circles show angle-averaged results from our sequence of models that differ only in the choice of ignition properties (which parametrizes the deflagration strength). Open circles show our models with differing WD central densities. Crosses indicate model N100 (which can be considered a part of both sequences of models). The vertical bars indicate the spread due to observer orientation.
{\it Top panel:} Peak absolute magnitudes in optical
bands ($U$, $B$, $V$, $R$ and $I$). {\it Middle panel:} Colours at $t_ {\rm max}^{B}$. The dashed lines 
indicate the typical observed peak colours for normal SNe~Ia (Blondin et al. 2011). {\it Bottom panel:} $B$-band light curve time scales (rise time to peak, $t_ {\rm max}^{B}$ and decline rate parameter $\Delta m_{15}^B$). The dashed line indicates the mean value of
$t_ {\rm max}^{B}$ derived from observations by \citep[][]{strovink07}. }
\label{fig:lctrends}
\end{figure}

As expected based on the $^{56}$Ni masses \citep[0.32 -- 1.1~M$_{\odot}$, ][]{seitenzahl13}, the models predict a wide range of brightnesses in the optical bands (Table~\ref{tab:lcs}, Figure~\ref{fig:lcs_1}), sufficient to encompass the range of spectroscopically normal SNe~Ia. In the models, this diversity is driven by the choice of the number and location of ignition sparks (which controls the strength of the deflagration phase): compared to the variations between models with different ignition conditions, the effect of varying the WD central density is relatively modest (Figure~\ref{fig:lcs_2}). 

The peak brightness variation between models is largest in the bluest bands (see top panel of Figure~\ref{fig:lctrends}): the models span a range in $U_{\rm max}$ of ${\simgt}2.5$~mag compared to only ${\sim}0.7$~mag in $I_{\rm max}$. Among models with the same WD central density, angle-averaged $U_{\rm max}$, $B_{\rm max}$ and $V_{\rm max}$ show nearly monotonic decrease (i.e.\ brightening) with increasing $^{56}$Ni mass. The trend in $R_{\rm max}$ and $I_{\rm max}$ is weaker and eventually turns over at the largest $^{56}$Ni masses. Our models with higher and lower central density are slightly displaced from the trends shown by the other models, but only by a few tenths of a magnitude.

The differing trends between photometric bands along our model sequence mean that the SN colours also vary. This is illustrated for $B - V$, $V - R$ and $V - I$ at $t_{\rm max}^{B}$ in the middle panel of Figure~\ref{fig:lctrends}. Each of these peak colours shows a ``brighter-bluer'' trend along our model sequence.
$B-V$ varies relatively little (and is always positive) but $V - I$ is more
than 1~mag bluer in our brightest models than in our faintest. As for the peak magnitudes in the blue bands, the 
colour changes roughly monotonically with the $^{56}$Ni mass among the models with the same WD central density.

For all the models, maximum light (in angle-averaged $B$-band) occurs between 16 and 18 days after explosion, roughly consistent with observations \citep{strovink07,hayden10,ganeshalingam11}. There is some systematic variation of $t_{\rm max}^{B}$ along our model sequence (see last panel of Figure~\ref{fig:lctrends}), but it is not very strong. The post-maximum light curve decline parameter ($\Delta m_{15}^B$) is also rather similar in all the models, typically around 1.3~mag. 

Comparison of our angle-averaged light curves to photometric observations of normal SNe~Ia clearly favours the models with intermediate deflagration strengths: N100H, N40, N100, N150, N100L and N300C.
In the $B$, $V$ and $R$ optical bands these models provide reasonable matches to the light curves of normal SNe~Ia (e.g.\ SN~2005cf [\citealt{pastorello07a}], shown in Figures~\ref{fig:lcs_1} and \ref{fig:lcs_2}). There are some significant shortcomings, however. In particular, for all the models, there is a systematic discrepancy in synthetic optical colours: the models are too red. For example, normal SNe~Ia at peak typically have $B-V$ in the range $-0.2$ to $0.0$ \citep[see e.g., ][]{blondin11}, 
while all our models have peak $B-V > 0$ at $t_{ \rm max}^{B}$ (see Figure~\ref{fig:lctrends} and Figure~\ref{fig:lc_obs_colours}).\nocite{hicken09} 
The redness problem is most severe for models with strong deflagrations ($\simgt 200$ ignition sparks) but it persists throughout the sequence, appearing as a clear offset in the relation between $B_{\rm max}$ and $V_{\rm max}$ as compared to observations (see Figure~\ref{fig:lc_obs_colours}). 
We attribute the red colours to excessive line blocking in the blue,
primarily by iron-group elements. Of particular importance in the current models is the location of the deflagration burning products. 
In 1D models (e.g.\ the W7 model of \citealt{Nomoto84} or the models of \citealt{hoeflich95}, \citealt{hoeflich96a} and \citealt{blondin13}), the products of highest density burning (e.g.\ stable iron-group material) are centrally concentrated but for models involving a deflagration this is an artificial consequence of the imposed spherical symmetry. Instead, owing to buoyancy, in our multi-dimensional simulations the deflagration ash is not confined to a low-velocity quasi-spherical core but rather forms a clumpy shell-like structure at intermediate velocities.
This material is exterior to most of the $^{56}$Ni-rich regions and immediately interior to (or even mixed through) the Si/S-rich regions in many of the models (see figures 3 and 4 of \citealt{seitenzahl13}). As such, it is able to significantly affect the spectrum in the photospheric phase around and after maximum light.

Our optical light curves tend to decline slightly more quickly post-maximum than do the bulk of normal SNe~Ia (the models typically have $\Delta m_{15}^B \sim 1.3$ and $\Delta m_{15}^V \sim 0.8$; see Figure~\ref{fig:lc_obs_WL}). More importantly, viewed as a sequence, the models fail to show clear width-luminosity relations (WLRs) in the optical bands, in contrast to observations (see Figure~\ref{fig:lc_obs_WL}). This suggests that the variation amongst our current models does not provide a good description of the differences between real events in the main SNe~Ia population, a key conclusion to which we will return in Section~{\ref{sect:discuss}}.

In the $I$-band the model light curves provide a rather poor match to observations, typically remaining too bright for too long. Accurate modelling of the $I$-band can be strongly affected by the Ca~{\sc ii} infrared triplet features. This becomes a prominent emission feature in the models post maximum light, the strength of which is likely overestimated since {\sc artis} does not currently account for forbidden line emission in iron-group elements. Consequently, the poor match in the $I$-band may be partially attributed to shortcomings in the radiative transfer calculations. We note, however, that better agreement was found using the same numerical codes for models of sub-Chandrasekhar mass explosions \citep{sim10} and merger models \citep{pakmor12} -- thus the $I$-band excess in the models may also point to a genuine shortcoming of the models.

In the NIR bands, the models with $>20$ ignition sparks qualitatively agree with observations: the NIR light curves generally show double peaks and the model-to-model variation in brightness around the earlier NIR peak is much less than in the optical bands. The sequence of models with moderate to strong deflagration phases are therefore consistent with the observation that the first NIR peak of normal SNe~Ia is a good standard candle \citep{elias85,meikle00,krisciunas04}. Overall, the first NIR peak is slightly too faint in the models but we note that this part of the light curve can be influenced by the completeness of the atomic data set \citep{kromer09} -- a more complete atomic data set would be expected to produce increased fluorescence to the NIR at early epochs.

\subsubsection{Orientation dependence of light curves}
\label{sect:lc-angle}

Thus far we have discussed only properties of our angle-averaged light curves. Since our models are fully 3D, however,
the synthetic observables also depend on observer orientation. 
The range of angle-dependence for a sample of our models is illustrated in Figure~\ref{fig:lcs_1} 
and the degree of variation caused by orientation is indicated for all the quantities plotted in Figure~\ref{fig:lctrends} and tabulated in Table~\ref{tab:lcs}. 
The orientation dependence of the peak magnitude in $V$, $R$ and $I$
is small (${\sim}0.1$~mag). In the $B$ and $U$ bands it is more
significant, typically ${\sim}0.2$~mag but increasing to $> 0.5$~mag
in some cases, particularly for models with weak deflagrations
(e.g.\ N3 and
N5)\footnote{We note that, despite having only one ignition spark,
  model N1 does not show the most pronounced viewing-angle
  effects. This is because the first detonation in model N1, which is necessarily triggered on the only deflagration plume present, occurred while {\it all} of the deflagration ash was still at relatively high density. The detonation can then completely surround the deflagration products, effectively interrupting the buoyant motion and leading to relatively spherical ejecta.
In contrast, in models N3 and N5, some deflagration plumes were close to reaching the surface when the first DDT took place, thereby imprinting strong asymmetries on all ejecta layers.}.
Nevertheless, for both peak magnitudes and colours, the angle-dependencies
are relatively modest compared to the mean trends along our model sequences.
In contrast, the light curve time-scale parameters $t_{\rm max}^{B}$ and $\Delta m_{15}^B$ both show orientation effects that have comparable amplitude to the differences between the models (as noted above, these parameters did not show such clear systematic trends along our model sequence). 

\begin{figure*}
\includegraphics[height=8.5cm, angle=90]{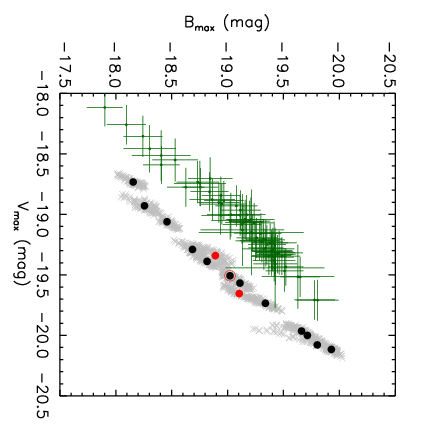}
\includegraphics[height=8.5cm, angle=90]{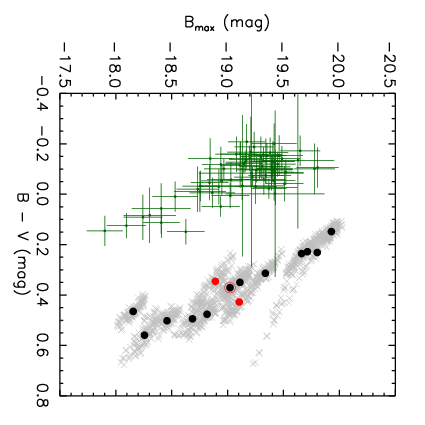}
\caption{Comparison of synthetic light curve properties between our models and a sample of SNe~Ia drawn from Hicken et al. (2009, excluding events with a distance modulus $\mu < 33$~mag.). {\it Left:} Peak $B$-band versus peak $V$-band magnitude. {\it Right:} $B$-band magnitude versus $B-V$ colour at $t_{\rm max}^B$. Observations are shown as green crosses while measurements from our angle-averaged model light curves are shown as filled circles [black for our twelve models that differ only in the number and distribution of ignition sparks; red for our two models that adopt different central densities; model N100 is shown as a black circle with a red ring, for comparison to both sequences]. The light grey crosses indicated results from the full orientation dependent synthetic light curves (100 points for each model).}
\label{fig:lc_obs_colours}
\end{figure*}

\begin{figure*}
\includegraphics[height=8.5cm, angle=90]{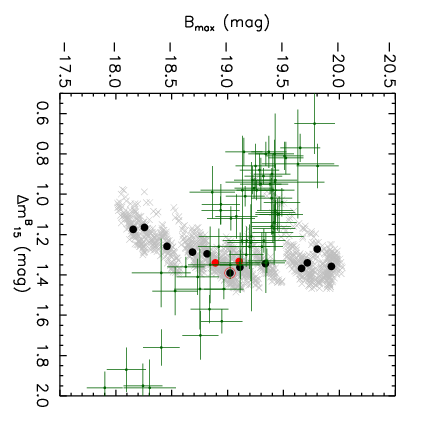}
\includegraphics[height=8.5cm, angle=90]{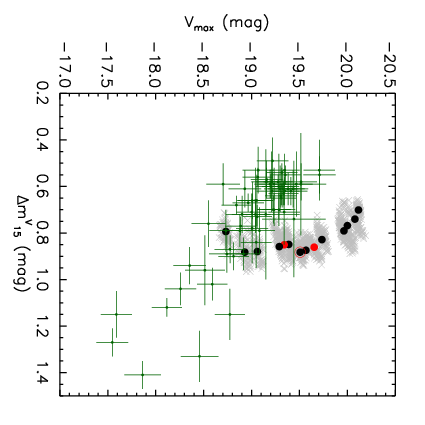}
\caption{As Figure 4 but showing $B$- (left) and $V$-band (right) width-luminosity relations. }
\label{fig:lc_obs_WL}
\end{figure*}

\begin{figure*}
\includegraphics[height=8.5cm, angle=90]{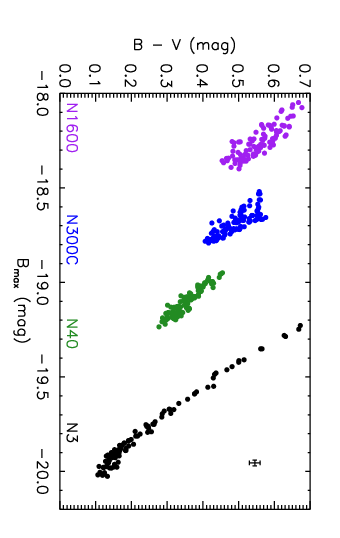}
\includegraphics[height=8.5cm, angle=90]{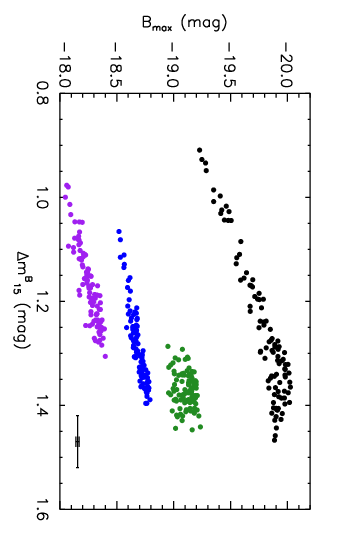}
\caption{Correlations between light curve properties for different
  observer orientations: $B - V$ colour at peak versus $B_{\rm max}$
  (left panel), $\Delta m_{15}^B$ versus $B_{\rm max}$ (right
  panel).
In each case we show results for four of our models (N3, N40, N300C and N1600; see left panel for colour coding). The error bars shown in black 
indicate the uncertainties in our calculations owing to a combination of Monte Carlo noise and the length of our simulation timesteps.}
\label{fig:lctrends2}
\end{figure*}

Some strong correlations are predicted amongst the orientation dependencies of the light curves, as illustrated in Figure~\ref{fig:lctrends2} (see also the line-of-sight differences in Figures~\ref{fig:lc_obs_colours} and \ref{fig:lc_obs_WL}). The peak $B$-band magnitude and $B-V$ colour at peak show a strong brighter-bluer relation for different lines of sight to each model. This is mostly driven by the viewing-angle variation of the $B$-band magnitude although a weaker brighter-bluer relation is still apparent when comparing $V_{\rm max}$ and peak $B - V$. 
The $B$-band light curve decline parameter, $\Delta m_{15}^B$, also shows a correlation with $B_{\rm max}$ (second panel of Figure~\ref{fig:lctrends2}) -- although there is some dispersion, viewing-angles associated with brighter $B_{\rm max}$ tend to have more rapid post-maximum decline. This orientation trend is in the {\it opposite} sense to the observed  WLR and will be discussed further in Section~\ref{sect:discuss} (see also Figure~{\ref{fig:lc_obs_WL}}). 

\subsection{Spectra}
\label{sect:spec}

We now turn our attention to comparing spectral features in our models to observations. In this section, we will carry out comparisons without reference to absolute brightness or broad-band colours -- comparisons of spectral features and light curve properties will be drawn together  in Section~\ref{sect:discuss}.

\subsubsection{Angle-averaged spectra}

\begin{figure*}
\includegraphics[height=11cm, angle=90]{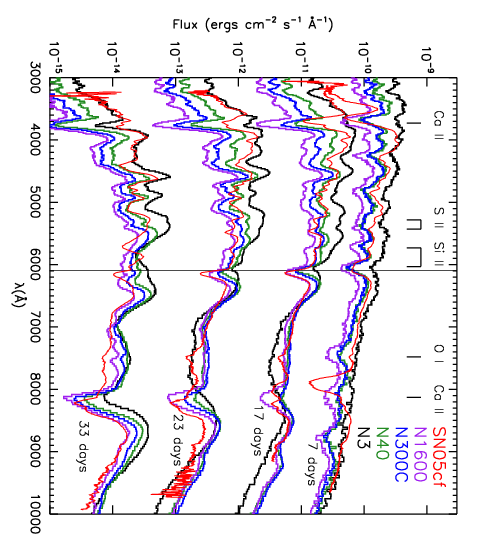}
\caption{Synthetic spectra for models N3, N40, N300C and N1600 for 7, 17, 23 and 33~days after 
explosion (distance of 1 Mpc). For clarity, the spectra have been shifted via multiplicative factors of [20 (for 7 days), 1 (for 17 days), 1/20 (for 23 days) and 1/400 (for 33 days)]. Positions of characteristic features due to intermediate mass elements in the synthetic spectra are marked. A
vertical line is drawn as a reference point for comparing velocities of the Si~{\sc ii} 6355\AA~line. For comparison, we also show spectra of the spectroscopically normal SNe~Ia 2005cf (Garavini et al. 2007; red lines) for epochs of -10, +0, +6.5 and +16.5~days relative to $B$-band maximum, roughly comparable with the epochs of the synthetic spectra shown. The observed spectra have been corrected for redshift and reddening (using values from Pastorello et al. 2007) and
scaled to account for the distance modulus and multiplicative factors applied to the model spectra.}
\label{fig:spec1}
\end{figure*}

In Figure~\ref{fig:spec1} we show angle-averaged synthetic spectra for four of our explosion models at four epochs (7, 17, 23 and 33~days after 
explosion).\nocite{garavini07}
In all our models, both the early (e.g.\ 7 days post explosion) and maximum light (roughly $17$~days post explosion) spectra show clear 
spectral features associated with intermediate mass elements (e.g.\ Si and S), as are characteristic of SNe~Ia. 

The strong Si~{\sc ii} 6355-\AA~line and the weaker Si~{\sc ii} 5972-\AA~line both show clear systematic variation along our model sequence. In the 7 and 17~day spectra, both lines are 
stronger in the fainter models. The variation in the Si~{\sc ii} 5972-\AA~line is greater, meaning that the ratio of the equivalent widths (EWs) of the two Si lines [EW(5972-\AA)/EW(6355-\AA)] clearly varies with luminosity, as is observed \citep{nugent95,hachinger08}. 
As time passes, the Si velocities decrease and the lines become weaker. In our 33~day spectra, the Si and S lines have largely disappeared -- the spectral formation is dominated by fluorescence in iron-group material, although there are still clear Ca features (including the H and K resonance features and the IR triplet). 

\begin{table*}
\caption{Parameters from measurements and fits to our synthetic spectra. 
MC noise and identification of continuum limit the accuracy of measuring $v_{\rm Si} (t_{\rm max}^{B})$ and $\dot{v}_{\rm Si}$ to around $\sim 300$~km~s$^{-1}$ and 50~km~s$^{-1}$~day$^{-1}$, respectively.
For the {\sc snid} matches we quote the SN name followed by the epoch of the observed spectrum (in days, relative to maximum light). In every case we give only the best matching (highest $r$lap) example. With the exception of the latest epoch considered for N1600C, all the best  {\sc snid} matches are to spectroscopically normal SNe~Ia (as classified in the {\sc snid} database).}
\begin{tabular}{lrrrcrrcrrcr}\hline
 & & & \multicolumn{3}{c}{$t = t_{\rm max}^{B} - 7$~days} & \multicolumn{3}{c}{$t = t_{\rm max}^{B}$} & \multicolumn{3}{c}{$t = t_{\rm max}^{B} + 7$~days}\\
Model & $v_{\rm Si} (t_{\rm max}^{B})$& $\dot{v}_{\rm Si}$$^a$ & $r{\rm lap}$ & {\sc snid} & $z$ & ~~~~~~$r{\rm lap}$ & {\sc snid} & $z$& ~~~~~~$r{\rm lap}$ & {\sc snid} & $z$ \\
 & ($10^3$ km s$^{-1}$) & (km s$^{-1}$ day$^{-1}$) & & match & & & match & & & match & \\
\hline
N1    & 15.9 &  $70$ & 11.7 &  94D($-9.2$) & $-0.009$ & 7.8 &  06X($+2.3$) & $-0.003$ & 5.6 & 96bo($+9.2$) & $-0.008$ \\
N3    & 14.7 & $110$ & 12.1 & 96bl($-4.0$) & $-0.008$ & 8.5 & 02hd($+1.4$) & $-0.009$ & 6.4 & 07bm($+9.6$) & $-0.009$ \\
N5    & 14.1 &  $70$ & 12.7 & 96bl($-4.0$) & $-0.005$ & 9.5 &  94T($+0.8$) & $-0.005$ & 6.9 & 04eo($+7.8$) & $-0.008$ \\
N10   & 14.0 &  $80$ & 14.1 & 96bl($-4.0$) & $-0.005$ & 8.9 &  94T($+0.8$) & $-0.005$ & 7.7 & 04eo($+7.8$) & $-0.007$ \\
N20   & 12.9 & $120$ & 15.4 & 07au($-4.7$) & $-0.006$ & 9.2 & 02hd($+1.4$) & $-0.004$ & 8.2 & 04eo($+7.8$) & $-0.005$ \\
N100H & 13.3 & $190$ & 17.2 & 02bo($-4.8$) & $-0.002$ & 7.6 & 04eo($+2.9$) & $-0.008$ & 9.1 & 07hj($+8.2$) & $-0.005$ \\
N40   & 12.9 & $190$ & 22.6 & 02bo($-4.8$) & $-0.002$ & 9.7 & 04eo($+2.9$) & $-0.007$ & 9.4 & 07hj($+8.2$) & $-0.003$ \\
N100  & 12.9 & $220$ & 21.8 & 02bo($-4.8$) & $-0.002$ & 9.8 & 04eo($+2.9$) & $-0.007$ & 9.5 & 07hj($+8.2$) & $-0.003$ \\
N150  & 12.5 & $190$ & 21.4 & 07au($-4.7$) & $-0.007$ & 7.8 & 04eo($+2.9$) & $-0.006$ & 8.9 & 07hj($+8.2$) & $-0.002$ \\
N100L & 12.4 & $200$ & 18.8 & 02er($-7.0$) & $-0.002$ &10.5 & 04eo($+2.9$) & $-0.006$ & 9.1 & 07hj($+8.2$) & $-0.002$ \\
N300C & 12.6 & $200$ & 21.7 & 07au($-5.6$) & $-0.006$ & 9.6 & 04eo($+2.9$) & $-0.006$ & 9.2 & 07hj($+8.2$) & $-0.002$ \\
N200  & 12.2 & $260$ & 19.4 & 02bo($-4.8$) & $-0.001$ &11.6 & 04eo($+2.9$) & $-0.005$ & 8.6 & 07hj($+8.2$) & $-0.000$ \\
N1600 & 11.9 & $260$ & 15.5 & 02er($-5.0$) & $-0.004$ & 9.4 & 04eo($+2.9$) & $-0.005$ & 8.7 & 07hj($+8.2$) &  $0.000$ \\
N1600C& 11.4 & $300$ & 14.5 & 08ar($-6.6$) &  $0.000$ & 9.1 & 04eo($+2.9$) & $-0.003$ &10.5 & 02es($+9.0$)$^b$ & $-0.009$ \\ \hline
\end{tabular}\\ 
$^a$ $\dot{v}_{\rm Si}$ parametrizes the rate of evolution of the Si~{\sc ii} 6355\AA~line at epochs around $t_{\rm max}^{B}$ (see Equation 1).\\
$^b$ 91bg-like
\label{tab:specs}
\end{table*}

\begin{figure}
\includegraphics[height=8.5cm, angle=90]{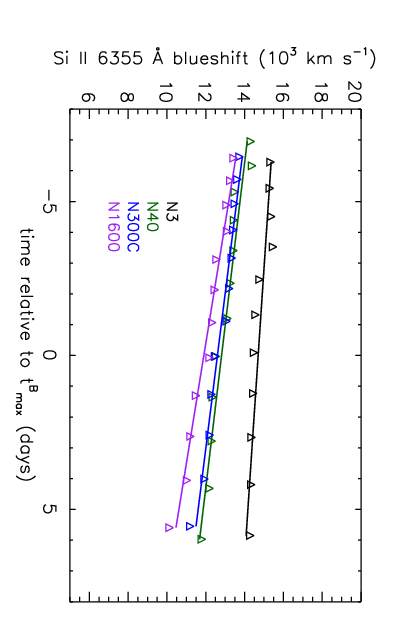}
\caption{Evolution of the blueshift of the Si~{\sc ii} 6355\AA~line for models N3, N40, N300C and N1600. Symbols show measurements from our synthetic spectra, which typically have an uncertainty of $\sim 300$~km~s$^{-1}$ due to MC noise and choice of continuum level. The solid lines show the simple linear fits used to characterise the rate of evolution at these epochs.}
\label{fig:si_evo}
\end{figure}

We have measured the blue-shift velocity ($v_{\rm Si}$) of the Si~{\sc ii} 6355-\AA~line at $t_{\rm max}^{B}$ for each model (given in Table~\ref{tab:specs}). In addition, we have also 
measured the line velocity for times between seven days before and six days after $t_{\rm max}^{B}$ (as noted above, the line generally fades from the spectrum soon thereafter). These measurements are shown for a sub-set of models in Figure~\ref{fig:si_evo}. Since the evolution is fairly simple, we have also fit a linear function to each model:
\begin{equation}
{v}_{\rm Si}(t) = v_{\rm Si} (t_{\rm max}^{B}) - \dot{v}_{\rm Si} \times (t -t_{\rm max}^{B}) \; \; . 
\end{equation}
Although simple, such fits provide a convenient rate-of-decrease parameter ($\dot{v}_{\rm Si}$) that can be compared between all the models.
The derived values of $\dot{v}_{\rm Si}$, which have a MC uncertainty of ${\sim}50$~km~s$^{-1}$~day$^{-1}$, are given in Table~\ref{tab:specs} and sample fits are shown in Figure~\ref{fig:si_evo}.

The range of blueshift $v_{\rm Si} (t_{\rm max}^{B})$ covered by our models ($11,400$ to $15,900$~km~s$^{-1}$) is roughly compatible with the maximum light velocities observed for normal SNe~Ia [${\sim}10,000$ to $14,000$~km~s$^{-1}$; see, e.g., figure 1 of \cite{benetti05a}; further discussion below]. As might be expected based on the correlation between kinetic energy and $^{56}$Ni mass (see Table~\ref{tab:lcs}),
the brighter models have higher $v_{\rm Si} (t_{\rm max}^{B})$. The mean rate of line-velocity evolution ($\dot{v}_{\rm Si}$) also varies systematically along the model sequence -- the velocity decreases most rapidly in the faintest models. 
Quantitative comparison between our $\dot{v}_{\rm Si}$-values and the velocity evolution parameters in 
\citet{benetti05a} is not easy because their measurements do not correspond to a fixed range of epochs. However, qualitative comparison consistently suggest 
that our synthetic spectra evolve too quickly:
in our spectra, the Si line is typically weak or absent within a week of maximum light while \cite{benetti05a} report many detections up to three or four weeks after $B$-band maximum (but see also \citealt{vanrossum13}).

Figure~\ref{fig:spec1} clearly illustrates verisimilitude between our synthetic spectra and observations of normal SNe~Ia (exemplified by SN2005cf). However, there are also clear discrepancies in the spectral features (e.g.\ around the Ca~{\sc ii} IR triplet). Consequently, it is interesting to consider
whether our model spectra would lead to classification as normal SNe~Ia or whether they provide a better match to one or other spectroscopically peculiar sub-class of SNe~Ia. To investigate this, we provide a 
rough classification of our synthetic spectra by  
using {\sc snid} \citep{blondin07}.

{\sc snid} is widely used in the classification of newly discovered SNe. It operates by first fitting and removing a pseudo-continuum from the spectra and then comparing the features in the flattened spectra to template SNe~Ia. This approach removes the sensitivity to absolute brightness and (largely) to colour, focussing instead on the spectral features. 
Consequently, some caution must be applied when interpreting the results (a good match in spectral features must be accompanied by agreement in light curve properties to constitute a good model). {\sc snid} comparisons nevertheless provide an excellent means to compare spectral features, the procedure by which observational classification is made.
{\sc snid} returns both a list of SN templates that are well-matched to the input spectrum along with a goodness-of-fit statistic ($r{\rm lap}$-value), which is based on the correlation between the input spectrum and the template (larger $r{\rm lap}$ values correspond to better matches; $r{\rm lap} > 5$ is considered a ``good'' match; \citealt{blondin07}). . The code also fits for a redshift parameter ($z$), based on comparison to the templates. When analysing models, this parameter can be viewed as a means to quantify any systematic velocity offset between features in the model and observations (see below). We made our comparisons to
a database of 3754 spectra of 349 template SNe~Ia (version 2.0 of the {\sc snid} template database, which includes data from \citealt{blondin12}).

\begin{figure}
\includegraphics[height=8.5cm, angle=90]{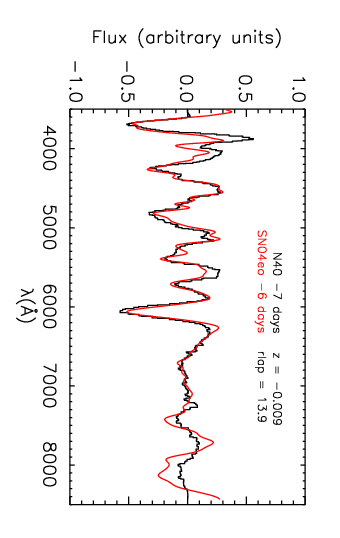}
\includegraphics[height=8.5cm, angle=90]{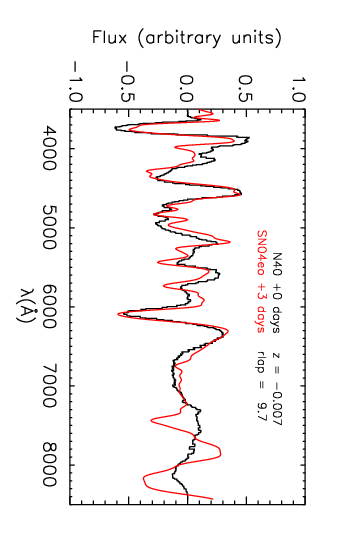}
\includegraphics[height=8.5cm, angle=90]{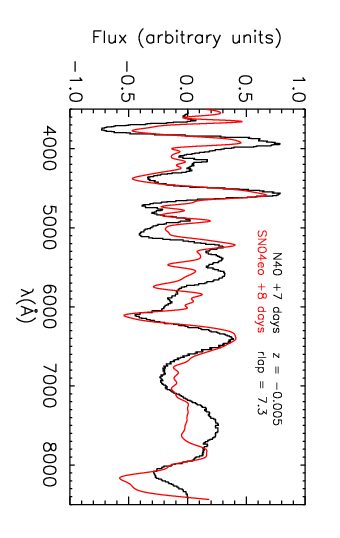}
\caption{{\sc snid} comparison of
spectral features in our N40 model to SN2004eo (the best match supernova identified for our N40 maximum light spectrum).
We show results for three epochs, $-7$, $0$ and $+7$~days relative to maximum light in $B$ band. 
The epoch of the relevant observation, the  $r{\rm lap}$-value of the match and the redshift parameter ($z$), is indicated in each panel.
The spectra are shown flattened, as they are compared by {\sc snid} (i.e.\ after a pseudo continuum has been fit and subtracted). Small red-shifts are allowed in identifying the best correlation. These are included here and can be used to quantify the typical mismatch in velocities between the models and observations (see text and Table 2).
}
\label{fig:snid}
\end{figure}

Following an approach similar to \cite{blondin11}, we ran {\sc snid} on angle-averaged synthetic spectra for three epochs for each of our models: at $t_{\rm max}^{B}$ and at both one week before and one week after $t_{\rm max}^{B}$. 
In all cases, we restricted the wavelength range of comparison to $3500 < \lambda($\AA$) < 8500$, forced the redshift associated with our synthetic spectra to be small and allowed comparison to template spectra for which the epoch (relative to maximum light) was within $\pm 3$~days of the synthetic spectrum.
Sample results from our {\sc snid} fits are presented in Table~\ref{tab:specs} and Figure~\ref{fig:snid}.
In all cases, our spectra yielded acceptable ($r{\rm lap} > 5$) matches to multiple SNe. For brevity, we tabulate only the best match SN in each case (see Table~1, which gives the SN name, epoch of the matched template, $r{\rm lap}$-value and redshift parameter for the best match to each synthetic spectrum).
Note that since each of our {\sc snid}  runs was carried out independently without restriction to any particular sub-set of the database, the best matching SN identified for a given model is not the same at all epochs considered. This is not particularly surprising -- from Section~\ref{sect:lcs}, it is clear that the time evolution of the models is imperfect. Moreover, it does not imply that insisting on comparison to one particular object leads to poor $r{\rm lap}$-values. This is illustrated for one of our models in Figure~\ref{fig:snid}, which shows {\sc snid} comparisons between N40 and SN2004eo at all three epochs analysed (we chose this comparison object since it is the best match at $t_{\rm max}^{B}$).

In all cases, at $t_{\rm max}^{B}$ and one week before $t_{\rm max}^{B}$, the best {\sc snid} 
match is with a template for a normal SNe~Ia (according to the classifications in the {\sc snid} database). 
Moreover, the quality of the fit is generally good ($r{\rm lap} > 7$ in all cases) and it is best for the models with moderate numbers of 
ignition sparks (N40, N100, N100L, N100H -- i.e.\ the same models which were most promising based on their light curve characteristics, see Section~\ref{sect:lcs}). 
We do note, however, that the models often fail to match well in the redder parts of the spectrum (around O~{\sc i} 7774~\AA\ and the Ca~{\sc ii} IR triplet, particularly at the later epochs).
Both for models with many fewer or many more ignition sparks, the best $r{\rm lap}$ value is generally poorer but still sufficiently high that, if realised in nature, such models would be spectroscopically classified as SNe~Ia.

At the later epoch (one week after $t_{\rm max}^{B}$), the best match is still with a normal SN~Ia for most models. At this epoch, the $r{\rm lap}$ value is typically slightly poorer although still good for our models with intermediate deflagration strengths (N40 -- N100).
For our very faintest model, the best post-maximum match is to a 91bg-like SN~Ia. This match with a peculiar SNe~Ia at the extreme of our model sequence is not particularly informative, however -- based on the light curves (i.e.\ absolute brightnesses), we can easily exclude our models with the largest numbers of ignition sparks from being appropriate for the bulk of normal SNe~Ia. Physical correspondence with 91bg-like SNe at the extreme of our model sequence is also unlikely based on light curves: our N1600 and N1600C models show slow post-maximum decline rates (the smallest $\Delta m_{15}^B$-values amongst our models), while observed 91bg-like SNe~Ia are noted for their rapid decline after maximum light. 

We note that for none of our comparisons was the best {\sc snid} match with a 1991T-like SN~Ia. Considering our five models with high $^{56}$Ni mass (N1, N3, N5, N10 and N20), 
the best $r{\rm lap}$-values in comparisons to 1991T-like events were only 7.4, 5.3 and 5.4 respectively for the $-7$, $0$ and $+7$ day epochs considered. These are substantially poorer than the matches found for spectroscopically normal SNe~Ia (see Table~\ref{tab:specs}), suggesting that our models are not particularly promising candidates to explain 1991T-like events.

Our {\sc snid} fits also allow us to more accurately assess the extent to which the line velocities that manifest in our spectra agree with observations. 
The best correlation is generally obtained when a small {\it blue}shift is applied to the template spectra, meaning that the velocities in the simulations are slightly too high (typically by around 1500~km~s$^{-1}$ for maximum light spectra). This systematic effect suggests that slightly too much kinetic energy is typically released. This might be a consequence of the assumed initial WD composition in our models. 
In particular, a smaller C/O ratio could lead to less energy generation and more slowly expanding ejecta -- this will be considered further in a forthcoming study.

\subsubsection{Orientation dependence of spectra}

\begin{figure*}
\includegraphics[height=8.5cm, angle=90]{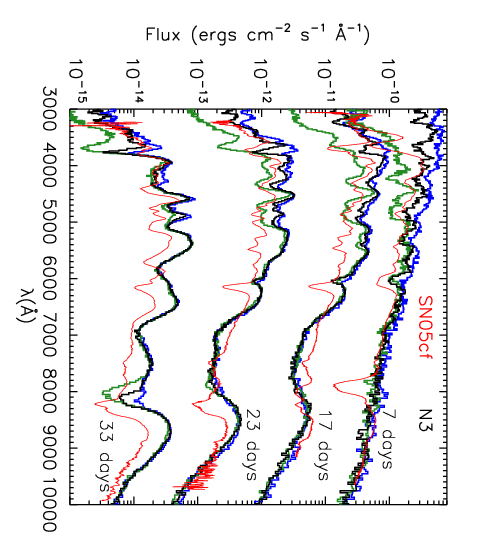}
\includegraphics[height=8.5cm, angle=90]{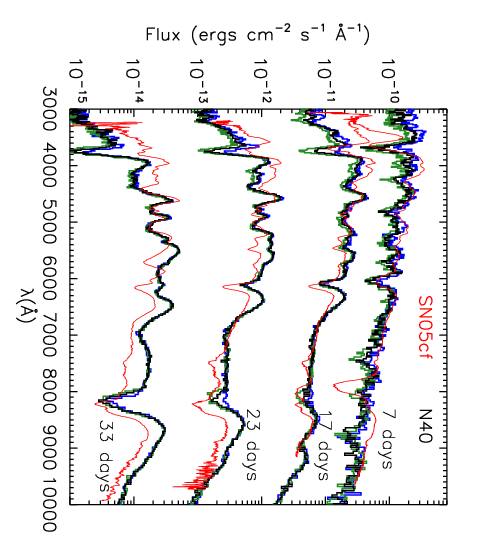}
\caption{Orientation-dependent spectra for models N3 and N40 for 7, 17, 23 and 33 days after explosion (distance of 1 Mpc). In black (blue/green) we show spectra representative of orientations with median (maximum/minimum) brightness. As in Figure 7, the spectra have been shifted via multiplicative factors for clarity and observations of SN~2005cf are shown in red.}
\label{fig:spec_res}
\end{figure*}

Figure~\ref{fig:spec_res} shows synthetic spectra for models N40 and N3 at four epochs for three example observer orientations, which correspond to the directions in which the peak bolometric magnitude was largest, smallest and close to the median value\footnote{In order to improve signal to noise in these MC spectra, each is obtained by averaging over 5 of the synthetic spectra drawn from our sample of 100 orientations: the 5 brightest, 5 faintest and 5 around median brightness.}.

As expected based on the discussion of light curves in Section~\ref{sect:lc-angle}, our synthetic spectra for model N40 are not very sensitive to the observer orientation (only the region around the Ca~{\sc ii} infrared triplet and the blue/ultraviolet region of the spectrum are noticeably affected). In particular, neither the strength nor blueshift of the Si~{\sc ii} $ 6355$-\AA~feature is significantly direction dependent. 
This degree of orientation dependence is characteristic of our models with moderate to large numbers of ignition sparks.

In contrast, the N3 spectral features do vary noticeably with orientation. At maximum light, the Si~{\sc ii} $6355$-\AA~ feature is less blueshifted (down to $v_{\rm Si} \sim 13,500$~km~s$^{-1}$) and weaker for orientations in which the total flux is higher. For the median brightness and brightest example spectra of model N3 around $t_{\rm max}^{B}$, the best {\sc snid} matches are still to normal SNe~Ia, as for the angle-averaged spectrum of this model. In contrast, the faint spectrum shows a strongest correlation with a SN~Ic (SN~04aw), having $r$lap = 7.5 (the best match with a SN~Ia has $r$lap = 5.7). However, we stress that the absolute brightness of model N3 (see Section~\ref{sect:lcs}) is too large for this to be a good model for a normal SN~Ia or SN~Ic. Therefore,
accounting for orientation dependence does not alter our conclusion that 
the brightest models do not appear well-matched to any observed class of SN (see below).

\section{Discussion}
\label{sect:discuss}

In the previous sections we have presented our synthetic light curves and spectra and compared them to the observed properties of
normal SNe~Ia. 
Here, we draw together the results of these comparisons and discuss their implications for the model parameters explored in our study. We also compare some key results to those obtained in previous studies of DDT models.

\subsection{Role of the strength of the deflagration phase}

Twelve of our fourteen models differ from each other only in the geometry with which the initial deflagration was ignited. These produce a sequence of models with differing deflagration strengths and final $^{56}$Ni masses ranging from 0.32 -- 1.1~M$_{\odot}$. Based on comparison of spectral features {\it alone}, almost all models in this sample invite comparison with normal SNe~Ia: apart from exceptions at the extremes of our model sequence, all our {\sc snid} analyses yielded best matches to normal SNe~Ia, with respectable $r$lap-values (particularly at early epochs). 

When we also consider light curve information, however, a clearer discrimination between our models emerges and our analyses consistently
point to the best agreement with observations for the models with intermediate strengths of the deflagration phase (e.g.\ N40, N100). These models have roughly the right rise time ($t^{B}_{\rm max}$), light curve decline parameter ($\Delta m_{15}^B$), peak luminosity and 
peak $V-R$ colour to match normal SNe~Ia.
The correspondence with observations is nevertheless imperfect -- for these models, important difficulties include matching the $I$ band light curve and the peak $B-V$ colours, which are systematically too red (see Section~\ref{sect:discuss-colour}). In addition, we consistently find that the models evolve too quickly post-maximum (e.g., although close, both the light curve decline parameters $\Delta m_{15}^B$ and $\Delta m_{15}^V$ are a little larger than in the bulk of normal SNe~Ia; see Figure~{\ref{fig:lc_obs_WL}}).

Although our models with weak deflagrations (i.e.\ those with very small numbers of ignition sparks: N1, N3, N5)  do have bluer $B-V$ colour at peak than our models with stronger deflagrations, they are still too red compared to the bulk of normal SNe~Ia (see Figure~\ref{fig:lc_obs_colours}). Especially in $V$-band, the N1, N3 and N5 light curves are somewhat too bright for most normal  SNe~Ia  and of particular note is the failure of these models to yield slower post-maximum evolution compared to their fainter counterparts (recall that normal SNe~Ia, and also 1991T-like SNe~Ia, are observed to obey a clear correlation between peak luminosity and light curve width); this suggests that our model sequence does not correctly capture the processes that drive systematically slower evolution in brighter objects (see Section~\ref{sect:discuss-WLR} below).
In addition, our models with few ignition sparks can be subject to very significant asymmetries that lead to orientation effects that are inconsistent with observations (see Section~\ref{sect:discuss-orientation} below). Taken together, these shortcomings lead us to disfavour our models with weak deflagration phases 
as promising matches to normal SNe~Ia. 
Furthermore, at present, we are not aware of any proposed sub-class of SNe~Ia that displays photometric properties similar to these models.

In comparison to normal SNe~Ia, our models with very strong deflagration phases fare somewhat better than those with weak deflagrations but are still inferior to the models around the middle of our sequence. In particular, N1600 and N1600C have $B_{\rm max}$-values that are clearly sub-luminous despite having $V_{\rm max}$ and $R_{\rm max}$ that are close to those of spectroscopically normal SNe~Ia. Despite being relatively faint, these models actually show the slowest post-maximum decline (smallest $\Delta m_{15}^B$) in our sequence. 
This is in sharp contrast to the sub-luminous 1991bg-like SNe, which typically display rapid post-maximum decline \citep[see e.g.\ ][]{taubenberger08} in addition to spectroscopic peculiarities.
Relatively gradual fading is observed in 2002cx-like SNe (sub-luminous SNe~Ia characterised by low expansion velocities; see e.g.\ \citealt{branch04,jha06}). However, as mentioned in Section 1, the best
models for 2002cx-like SNe appear to be pure deflagration events \citep{branch04,jha06,phillips07,jordan12b,kromer13}. 
We note that it was to be expected that our simulations with large numbers of ignition sparks bear some resemblance to pure deflagration models (e.g.\ very red colours and relatively slow post-maximum decline): 
as the number of ignition sparks is increased in our DDT models,
the deflagration phase becomes increasingly dominant in the explosion.

\subsection{Role of central density}

Three of our models (N100, N100H and N100L) differ from each other only in the choice of the central density of the WD. 
As would be expected from the differences in $^{56}$Ni mass, the peak brightnesses of these models are only moderately different from each other (spanning a range of only ${\sim}0.3$~mag in $V$ band). Therefore, for our model sequence, we conclude that the influence of the central density on the observables is of secondary importance to deflagration strength.
Nevertheless, the effect of central density is potentially observable and it could act as a secondary parameter \citep[e.g.\ ][]{hoeflich10,krueger11,seitenzahl11,krueger12,calder13}. We do note that, in our simulations, there is a weak but systematic trend for redder $B-V$ colour with increasing central density, a consequence of the larger mass of iron-group elements produced. Thus, variations in central density could drive dispersion in e.g.\ the colours of SNe~Ia produced via DDT events.

\subsection{Orientation effects}
\label{sect:discuss-orientation}

The light curves for all our models show some sensitivity to the observer inclination angle: in particular, we obtain more rapid post-maximum decline for orientations in which the explosion is seen to be brighter. This leads to a predicted anti-correlation between $\Delta m_{15}^B$ and $B_{\rm max}$ for the same explosion viewed from different directions. Such a correlation is {\it opposite} to the sense of the observed WLR for normal SNe~Ia (but characteristically similar to that due to orientation effects in other asymmetric models; see e.g.\ \citealt{sim07b}). Consequently, orientation effects in these models cannot play a role in explaining the primary trend of the SN~Ia WLR.

In our most asymmetric model (N3), the scale of the orientation effect is so large that it can be used to effectively exclude the model as being viable for normal SNe~Ia. However, in the other models (particularly those with modest to large numbers of ignition sparks), the effects of orientation are comparable to the scatter about the observed WLR and may therefore provide a satisfactory explanation for the observed spread around the mean WLR.

\subsection{The light curve width luminosity relation in our models}
\label{sect:discuss-WLR}

Viewed as a sequence, our 
DDT simulations do not show a clear WLR, in contrast to observations of normal SNe~Ia (see Figure~\ref{fig:lc_obs_WL}).
In particular, although they span a range of $B_{\rm max}$ and $V_{\rm
  max}$ that is wide enough to cover the whole range of observed
luminosities for normal SNe~Ia, all our models have very similar
post-maximum decline rate parameters $\Delta m_{15}^B \sim
1.3$~mag. The implication of this is that varying the strength of the deflagration {\it alone} (as parametrized by the ignition configuration)
does not necessarily
lead to a sequence of models that populate the light curve parameter
space occupied by normal SNe~Ia. 

Previous theoretical studies of sets of one- and two-dimensional DDT models and associated synthetic observables have produced WLRs that are roughly consistent with observations. For example, \cite{hoeflich96a} show that their 1D delayed detonation models follow a width luminosity relation. However, along their sequence of models the variation is not in the ignition conditions but in their choice for when the transition to detonation occurs (parametrized in terms of a transition density). This is also true for the 1D models recently analysed by \cite{blondin13} but is quite 
different from our sequence of 3D models in which the DDT criterion is not tuned (i.e the parameters that enter our model for the DDT mechanism are held fixed),
but rather the parametrized ignition configuration is used to control the deflagration strength. 

In their set of 2D models, \cite{kasen09} varied parameters for both the ignition configuration
and the DDT criterion. Careful analysis by
\cite{blondin11} showed that many of the \cite{kasen09} models (excluding very asymmetric setups) provide good matches to SNe~Ia and that the set of models populates the space of the observed SN~Ia WLR reasonably well (e.g.\ figure 7 of Blondin et al. 2011). 
However, we note that their models do show a correlation between the DDT parameter used and the light curve width parameter $\Delta m_{15}^B$. In particular, while their models with
their fourth and fifth DDT criteria (dc4 and dc5 sequences; \citealt{kasen09,blondin11})  have, on average, $\Delta m_{15}^B \sim 1.3$ (i.e.\ similar to our models), 
only their models in the dc1 and dc2 sequences typically yield $\Delta m_{15}^B \simlt 1$, which is characteristic of the observed SNe~Ia that are missed by our calculations.

We therefore conclude that although varying the rate with which material is consumed during the deflagration phase has a large effect in DDT models, changing this alone does not produce a sequence of simulations that 
accounts for the observed variation amongst SNe~Ia. Our results suggest that either (i) the deflagration strength is similar in most cases (being well-represented by our models with moderate numbers of ignition sparks) and the variations from object to object are controlled by an unrelated effect (e.g.\ stochasticity in the DDT process)
(ii) some additional explosion property (e.g.\ progenitor metallicity or properties of the DDT mechanism) varies in a way that is strongly correlated with the strength of the deflagration (note that all parameters of our models were held fixed in this study {\it except} the deflagration ignition configuration and the WD central density);   
or (iii) Chandrasekhar-mass DDT models are not responsible for the bulk of normal SNe~Ia. 

\subsection{Peak colours in our models}
\label{sect:discuss-colour}

As noted in Section~\ref{sect:lcs} and illustrated in Figure~\ref{fig:lc_obs_colours}, our 3D DDT models are systematically too red compared to observations of normal SNe~Ia (by around ${\sim}0.4$~mag in $B-V$ at $t_{\rm max}^{B}$). We attribute this to the location of iron-group elements (particularly the deflagration ash) in intermediate layers of the ejecta and we note that this occurs even for our models with central and near-symmetric ignition (such as N300C) -- in none of the models was the deflagration ash prevented from floating towards the surface \citep{seitenzahl13}.

The \cite{kasen09} 2D models were also noted to be systematically too red, in that case by ${\sim}0.2$~mag in $B - V$ \citep{blondin11}. The problem of red colours is less significant in studies of 1D models. For example, \cite{hoeflich96b} argue that, although slightly too red, their 1D DDT models yield $B-V$ colours that are consistent with observations, given uncertainties in the modelling. Similarly, the recent study of \cite{blondin13} also gives peak colours (they focus on $B - R$) that are within the range observed for normal SNe~Ia (in contrast, as for $B-V$, our 3D models with moderate numbers of ignition sparks give $B-R$ at peak that is too red by ${\sim}0.3 - 0.5$~mag). As noted by \cite{blondin13}, direct comparison is complicated here since the 1D simulations can include more sophisticated radiative transfer and ionization calculations than do our multi-dimensional models: non-LTE effects can influence the colours by a few tenths of a magnitude \citep[][]{hoeflich96b,kasen09}. Nevertheless, we do note that, even with our radiative transfer methods, we have obtained bluer colours for models that do not involve a deflagration phase 
[see e.g.\ \cite{sim10}, \cite{pakmor12} and \cite{roepke12}: in all those cases, models with peak $B-V \sim 0.2$~mag were found for luminosities corresponding to normal SNe~Ia]
 and for the 1D Chandrasekhar mass W7 model, in which the stable iron-group elements are present but artificially restricted to low velocities \citep{kromer09}. Thus our modelling methods certainly predict that the differential effect of deflagration ash in 3D models is to induce redder colours, which may be a major challenge. 
Nevertheless, we acknowledge that the scale of this effect may be comparable to uncertainties introduced by assumptions in the modelling and further investigation is warranted.

\section{Summary, implications and future work}
\label{sect:summary}

We have presented synthetic light curves and spectra for a sequence of 3D DDT models for SNe~Ia. Study of 3D effects is particularly important for DDT models owing to the complexity of modelling the buoyancy-unstable deflagration phase.
Our models differ from each other in two important respects, both of which are therefore probed by our study: the strength of the deflagration phase (as parametrized by the number and location of the ignition sparks) and the central density of the WD star. 

As described by \cite{seitenzahl13}, the model sequence gives rise to explosions with a sufficiently wide range of $^{56}$Ni masses to encompass the range covered by normal SNe~Ia. This is confirmed by the range of brightnesses of our synthetic light curves, which also have rise times compatible with observations. Moreover, inspection of our synthetic spectra around maximum light shows a fairly good match to the observed spectral features of normal SNe~Ia. Comparing our models clearly shows that variation in the strength of the deflagration flame can have a greater influence on the explosion and its synthetic observables than does the WD central density (which is varied by a factor of ${\sim}5$ in our study).

To interpret the results obtained from comparison of our models to observations, we return to the two purposes of our set of models (see Section~\ref{sect:models}): (i) to investigate whether good agreement with properties of a normal SNe~Ia can be obtained for some choice of the parametrized ignition in our simulations and (ii) to explore whether variations in the deflagration strength can account for (a part of) the observed diversity of SNe~Ia. 
With regard to (i), it is clear that 
the best match between any of our models and observations consistently occurs around the mid point of our sample (models N40, N100) and that, in many respect, such models fare rather well. In particular, these models show the best spectral match to normal SNe~Ia (as parametrized by the {\sc snid} $r$lap values) and have light curve rise times, peak brightnesses and decline timescales comparable to the bulk of normal SNe~Ia. In addition, these models are sufficiently close to spherical symmetry that line-of-sight effects are modest -- orientation effects in these models could explain the scatter across the observed WLR but do not lead us to reject the models.
However, on detailed inspection, several shortcomings remain. 
In particular, our models predict line velocities that are systematically too high (by ${\sim}1500$~km s$^{-1}$ for Si~{\sc ii} at maximum light) and colours that are too red (by around 0.4~mag in $B - V$ at maximum light). While exploring initial WD models of differing composition may provide a solution to the velocity mismatch, we suggest that the red colours are a consequence of the distribution of iron-group elements in the models (in particular, the deflagration ash), which may be a significant concern. Further modelling of this is certainly warranted, including studies that fully explore the possible geometry and location of the deflagration ash in 3D simulations and that properly consider the sensitivity of conclusions to approximations made in the radiative transfer (including a more sophisticated treatment of non-LTE effects).

The implication of our models for point (ii) are clearer.
The level of agreement between our synthetic observables and normal SNe~Ia becomes poorer at both ends of our model sequence. Moreover, our models show little systematic variation in light curve decline rates along the model sequence (Figure~\ref{fig:lc_obs_WL}).
Thus, viewed as a complete set, our models do not reproduce the observed SN~Ia WLR.
While uncertainties in the modelling may account for systematic shortcomings in the models (e.g.\ absolute mismatches in colours or velocities as mentioned above), it is to be expected that predictions for differential effects along the model sequence are more robust.  
Consequently, we suggest that the failure of our models to predict any strong WLR  
indicates that either the rate of fuel consumption during the deflagration phase is not the primary parameter driving the variation between explosions in a population of SNe~Ia produced via the DDT Chandrasekhar-mass scenario, or that the observed population of SNe~Ia is not dominated by events from this explosion channel. Further investigation of the sensitivity of 3D simulations to other aspects of the modelling, e.g.\ the choice of DDT criterion (as has been done in 1D and 2D), are needed to explore this further.

It is noteworthy that our models with few ignition sparks (N1, N3) fare most poorly in comparison with observations of SNe~Ia (they are typically too bright and can show strong orientation effects that are inconsistent with the observed tightness of the WLR). 
As noted in Section~\ref{sect:models}, our calculations are not based on any particular ignition simulations for SNe~Ia. However,
among our models, N1 and N3 are likely the best representations of explosions with single-spot off-centre ignition (in contrast, our models with larger numbers of sparks are more likely appropriate for centrally ignited explosions; see Seitenzahl et al. 2013).
In light of recent studies that favour ignition of the deflagration at an off-centre point \citep{zingale09,nonaka12}, the shortcomings of our models with few sparks therefore pose a particular challenge to the Chandrasekhar-mass DDT scenario for normal SNe Ia.
This lends credence to the alternative scenario in which 
off-centre ignition of a Chandrasekhar mass WD leads to explosion via pure deflagration (i.e.\ no DDT occurs), 
giving rise to faint spectroscopically-peculiar transients rather than normal SNe~Ia \citep{kromer13}.

Even if Chandrasekhar-mass DDT models are not
appropriate for the bulk of normal SNe~Ia, such explosions might still
be present in the population, and our models can therefore be used to
guide their identification. For example, \cite{foley12a} provided
evidence that SNe~Ia with circumstellar material (detected via
blue-shifted Na~{\sc i} absorption) have preferentially higher Si~{\sc
  ii} velocities and redder colours than the average SN~Ia. Here, a tantalising
 correspondence arises since, as noted above, our models do manifest both
high line velocities and red colours. Also, our
models are appropriate for the Chandrasekhar-mass single-degenerate
scenario, the channel for which it has been claimed that blue-shifted
Na~{\sc i} absorption can be most readily explained (see
e.g.\ \citealt{patat07,sternberg11}, but see also
\citealt{shen13,soker13}).
However, further study is needed and we caution that both the line velocities and colours may be subject to significant systematic uncertainties due to assumptions of the modelling. 

Finally, we reiterate the continuing need for improving the accuracy of modelling. As noted by e.g.\ \cite{blondin13}, currently a trade off is often made between 1D studies (in which spherical symmetry is imposed as an ad hoc assumption; e.g.\ \citealt{hoeflich96b,blondin13}) and multi-D studies (in which cruder approximations in the radiative transfer are made necessary by considerations of computational expediency: e.g.\ \citealt{kasen06a}, \citealt{kromer09}, \citealt{kasen09}, this study).
It is clear, however, that the agreement between several explosion models (including Chandrasekhar-mass DDT models) and observations of SNe~Ia is sufficiently close that detailed evaluation can be limited by the accuracy of modelling assumptions \citep[see e.g.\ ][]{roepke12}. For example, realistic modelling of departures from LTE is important in SNe~Ia, particularly during the post-maximum decline phase (e.g., compared to the typical error associated with measurements of $\Delta m_{15}^B$, the $B$-band light curve evolution is quite sensitive to the non-LTE treatment of ionization; see e. g. \citealt{kromer09}). Such modelling sensitivity must be borne in mind when comparing calculations to data. At the same time, 
multi-D studies show that asymmetries in realistic explosion models affect synthetic light curves and spectra by observable amounts, particularly (but not exclusively) in the blue and ultraviolet regions (e.g.\ the $\Delta m_{15}^B$-spread due to orientation effects in our models is comparable to the observed dispersion about the SNe~Ia WLR). 
Consequently, if theoretical predictions are to match the quality of data provided by current SN~Ia observational programs, there remains a clear need for ongoing development of modelling tools that should ultimately include both the best possible micro-physics and handle the most realistic, multi-dimensional explosion models.

\section*{Acknowledgments}

Parts of this research were conducted by the Australian Research Council Centre of Excellence for All-sky Astrophysics (CAASTRO), through project number CE110001020.
This work was supported by the Partner Time Allocation (Australian National University), the National Computational Merit Allocation and the Flagship Allocation Schemes of the NCI National Facility at the Australian National University, by the Deutsche Forschungs\-gemeinschaft via
the Transregional Collaborative Research Center TRR 33 ``The Dark
Universe'', by the Excellence Cluster EXC153 ``Origin and Structure
of the Universe'', the graduate school ``Theoretical Astrophysics and Particle Physics'' at the University of W\"urzburg (GRK1147), by the Emmy Noether Program (RO 3676/1-1), by the ARCHES prize of the German Ministry of Education and Research
and by the Group-of-eight/Deutscher Akademischer Austausch Dienst Joint Research Co-operation Scheme. Parts of the 
simulations were carried out at the John von Neumann Institute for 
Computing (NIC) in J\"{u}lich, Germany (project hmu13) and within the Partnership for Advanced Computing in Europe (PR042).

\bibliographystyle{mn2e}
\bibliography{snoc}

\label{lastpage}

\end{document}